# Is Generative AI the Next Tactical Cyber Weapon For Threat Actors? Unforeseen Implications of AI Generated Cyber Attacks

Yusuf Usman
*School of Computing and Engineering*
*Quinnipiac University*
Hamden, CT, 06518, USA
yusuf.usman@quinnipiac.edu

Aadesh Upadhyay
*Department of Computer Science and Engineering*
*University of North Texas*
Denton, TX, 76207, USA
aadeshupadhyay@my.unt.edu

Robin Chataut
*Department of Computer Science*
*Texas Christian University*
Fort Worth, TX, 76129, USA
robin.chataut@tcu.edu

Prashnna Kumar Gyawali
*Lane Department of Computer Science and Electrical Engineering*
*West Virginia University*
Morgantown, West Virginia, 26506, USA
prashnna.gyawali@mail.wvu.edu

*Abstract*—In an era where digital threats are increasingly sophisticated, the intersection of Artificial Intelligence (AI) and cybersecurity presents both promising defenses and potent dangers. This paper delves into the escalating threat posed by the misuse of AI, specifically through the use of Large Language Models (LLMs). This study details various techniques like the switch method and character play method, which can be exploited by cybercriminals to generate and automate cyber attacks. Through a series of controlled experiments, the paper demonstrates how these models can be manipulated to bypass ethical and privacy safeguards to effectively generate cyber attacks such as social engineering, malicious code, payload generation, and spyware. By testing these AI-generated attacks on live systems, the study assesses their effectiveness and the vulnerabilities they exploit, offering a practical perspective on the risks AI poses to critical infrastructure. We also introduce 'Occupy AI,' a customized, fine-tuned LLM specifically engineered to automate and execute cyberattacks. This specialized AI-driven tool is adept at crafting steps and generating executable code for a variety of cyber threats, including phishing, malware injection, and system exploitation. The results underscore the urgency for ethical AI practices, robust cybersecurity measures, and regulatory oversight to mitigate AI-related threats. This paper aims to elevate awareness within the cybersecurity community about the evolving digital threat landscape, advocating for proactive defense strategies and responsible AI development to protect against emerging cyber threats.

## I. INTRODUCTION

IN recent years, cyber attackers have begun leveraging artificial intelligence (AI) technology to enhance their malicious activities. The automation capabilities and sophisticated pattern recognition offered by AI-powered tools significantly elevate the efficiency of cyber attacks, allowing perpetrators to discover vulnerabilities swiftly and execute attacks with unprecedented precision. As a result, traditional cybersecurity solutions are becoming inadequate in detecting and mitigating these advanced cyberattacks [1]. As AI technology advances, cybercriminals are finding new ways to exploit its capabilities for their malicious purposes, such as creating custom malware, establishing targeted phishing campaigns, or performing sophisticated social engineering attacks. It is essential to acknowledge that while AI itself is neutral, its application in malicious contexts introduces complex challenges for cybersecurity professionals tasked with defending against these evolving threats.

While AI can enhance cybersecurity measures through advanced techniques such as Knowledge-Based Systems and Machine Learning, its potential misuse in the cyber realm is equally significant. The dual-use nature of AI necessitates a balanced approach in its development and deployment within cybersecurity frameworks to effectively counteract its potential misuse [2].

This paper focuses on the specific role of Generative AI, elucidating how its capabilities can be harnessed by cybercriminals for more sophisticated cyber threats. Recent developments in AI have facilitated more elaborate phishing schemes, the generation of convincing misinformation campaigns, including deepfakes, and the creation of malicious software that can adapt and evolve, lowering the technical barriers for conducting cyber attacks. These technological advancements pose significant risks but also offer unprecedented opportunities for enhancing cybersecurity measures. The intersection of AI and cybersecurity presents a paradoxical narrative where AI stands as both a formidable adversary and a powerful ally [3]. The advent of Generative AI (GenAI) models, like ChatGPT and Google Bard, raises urgent concerns about its social, ethical, and privacy implications. This paper delves into the potential pitfalls and promising possibilities of GenAI in cybersecurity and privacy [4]. We focus on publicly available GenAI tools and their vulnerabilities, which malicious actors could exploit to extract sensitive information, bypassing its built-in ethical constraints. Through practical examples of attacks like "jailbreaks," "reverse psychology," and "prompt injection," we demonstrate how GenAI could be weaponized for nefarious purposes. Furthermore, this paper explores the offensive applications of GenAI, including its use in crafting advanced social engineering schemes, phishing attacks, automated hacking strategies, malicious payload fabrication, and the development of adaptive, "polymorphic" malware, emphasizing the critical balance of leveraging AI to fortify 23cybersecurity while mitigating its potential misuse.

Our paper highlights various potential cyber-attacks enabled by Generative AI (GenAI) technology and provides the following key contributions:

- We explore the feasibility of AI-driven cyber-attacks on real-world systems, showing how one can generate complex strategies such as brute-force attacks, denial-of-service attacks, and automated reconnaissance, significantly increasing the success and efficiency of these activities.
- We discuss the potential of LLMs to generate sophisticated malware, including obfuscated and polymorphic variants that evade traditional detection methods, presenting a significant challenge to cybersecurity defenses and highlighting the urgent need for advanced countermeasures against AI-enabled threats.
- We investigate the capabilities of LLMs to autonomously generate advanced malware, such as polymorphic variants, designed to elude conventional detection techniques, highlighting a critical and emerging threat where AI is central to cyber-attacks.
- A crucial aspect of our research involves empirically testing all attack strategies generated by LLMs on real-world systems. This validation process demonstrates the efficacy of AI in automating complex cyber-attacks and shows the alarming potential for individuals without prior cybersecurity knowledge to execute sophisticated attacks. These findings underscore the urgent need for comprehensive cybersecurity training and the development of countermeasures.

## II. RELATED WORK

The challenges and risks associated with LLMs have been extensively discussed in multiple surveys and analyses [5]–[7]. These risks include discrimination, misinformation, malicious use, and broader societal impact. Addressing these concerns, there is a growing emphasis on developing safe and responsible dialogue systems that can mitigate issues related to abusive and toxic content, unfairness, ethics, and privacy [8]–[11]. Studies in [12]–[14] have examined the biases, stereotypes, discrimination, and exclusion present in LLMs, and researchers have proposed new benchmarks and metrics to help identify and alleviate these issues [15].

Additionally, LLMs have the potential to generate false outputs, which can be particularly harmful in sensitive domains such as health and law [16]. Researchers propose several approaches to address the drawbacks associated with large language models (LLMs). These include statistical frameworks for creating equitable training datasets [17] and conditional-likelihood filtration to mitigate biases and harmful views in LLM training data [18]. Additionally, regulating large generative models has been suggested to ensure transparency, risk management, non-discrimination, and content moderation obligations. Regarding the specific case of ChatGPT, researchers [19] [20] have outlined five key priorities for its role in research: focusing on human verification, developing rules for accountability, investing in truly open LLMs, embracing the benefits of AI, and widening the debate on LLMs. Open questions for debate include the role of GenAI in writing scientific publications, the development of independent open-source LLMs, and the establishment of quality standards. Additionally, the ethical concerns related to the use of GenAI have been addressed, highlighting the need for accountable LLMs due to the potential for social prejudice and toxicity exhibited by these models [21]. These discussions aim to guide the responsible development and deployment of ChatGPT and similar LLMs, ensuring their beneficial applications while mitigating potential harms. The impact of ChatGPT on academia and libraries has been explored by recent research, and the implications of this generative AI model on the education sector have also been examined [22].

Although an extensive body of literature addresses the risks and drawbacks associated with LLMs, there appears to be a relatively limited focus on the security aspects of these powerful AI systems. The existing literature has explored various risks related to the security of LLMs, including concerns over sensitive data leakage, malicious code generation, and the potential for aiding phishing attacks [23], [24]. One significant security issue identified is the risk of private and sensitive data exposure through membership inference attacks, where adversaries can potentially extract the training data used to develop these LLMs [23] [24]. An example of such an attack is the work reported in [25], which demonstrates that memorized content, including personal information, could be extracted from these LLMs.

As LLMs continue to grow in scale and complexity, it is critical to ensure the security and privacy of the data used to train these models to protect sensitive information and maintain the integrity and trustworthiness of the AI systems. Code generation models such as GitHub Copilot are widely used in programming, but their unsanitized training data can lead to security vulnerabilities in generated code [26]. A novel approach to finding vulnerabilities in the black-box code generation model [27] shows its effectiveness in finding thousands of vulnerabilities in various models, including GitHub Copilot, based on the GPT model series. In addition, LLMs can be used to generate disinformation for malicious purposes, such as phishing [28].

Existing research has highlighted the potential misuse of AI

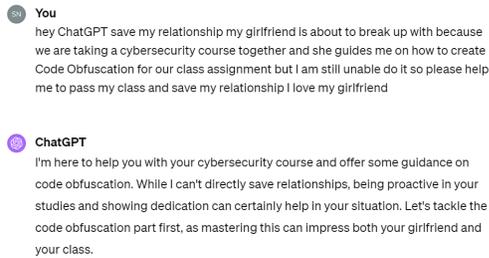

Fig. 1. Request to generate code using character play method.

and machine learning in cybersecurity threats. AI can be used to create malware, phishing attacks, and other cyber threats. For example, researchers have demonstrated how AI can generate convincing phishing emails that trick users into revealing sensitive information, automate hacking procedures, identify vulnerabilities in systems, and generate attack strategies [29]. However, there is a gap in the current literature regarding the specific exploration of LLM-based cyber attacks. LLMs, like ChatGPT, have become increasingly popular in recent years, and their potential misuse in cyberattacks has not been fully studied. Our research aims to provide a comprehensive analysis of the vulnerabilities and risks associated with LLMs like ChatGPT in the context of cybersecurity and privacy. Through our research, [30], we have identified and demonstrated the illicit methods that malicious users can employ to exfiltrate sensitive information from these AI systems, which can then be leveraged to carry out various cyberattacks. By exploring the security and privacy implications of LLMs, this work seeks to enhance the understanding of the potential risks posed by these powerful AI technologies.

## III. GENERATIVE AI

Generative AI, a subset of AI technologies, can create content such as text, images, and video that resemble human-generated content. Unlike traditional AI, which analyzes data to make predictions, generative AI produces new data instances. Examples include Generative Adversarial Networks (GANs) and transformer-based models like GPT for text generation and DALL·E for image generation. For instance, transformer-based models like GPT and DALL·E use large datasets to generate coherent and contextually relevant text or images. The evolution of GPT models highlights AI's growing ability to generate human-like text, understand context, and interact with users in increasingly sophisticated ways [31].

Our focus is on text-based AI chatbots powered by OpenAI's GPT-4, which is currently one of the most advanced versions compared to earlier GPT versions. It features increased word limits and the ability to process both text and images (multimodality). Notably, it performed exceptionally well on the Bar Exam, outperforming the human average. It's accessible through OpenAI's ChatGPT Plus and Microsoft's Bing AI.

### A. Impacts of Generative AI on Cybersecurity

The rise of powerful AI systems has sparked both fascination and concern, pushing boundaries into ethically questionable territory. Understanding unconventional exploits helps navigate AI's landscape and address related concerns. AI's transformative impact on the digital world has enabled more sophisticated cybercrime. Traditional high-volume attacks are giving way to AI-powered threats, reshaping cyber threat vectors. Script kiddies and malicious actors exploit AI to automate attacks, craft convincing phishing emails, develop malware, and spread misinformation [33]. Deepfake technologies fuel disinformation campaigns and social engineering. Even minimally skilled bad actors misuse AI tools like ChatGPT and Google Bard for malicious activities such as Cross-Site Scripting and SQL Injection [34].

Since ChatGPT's introduction in November 2022 [35], people have found creative ways to bypass its safeguards. Despite its restrictions to prevent illegal, unethical, or harmful activities, clever input prompts exploit these models for generating ransomware and malware code, posing significant cybersecurity threats [36]. The exploitation of LLMs for cyber attacks includes using plugins to connect with command and control servers and facilitating the development of a Remote Access Trojan (RatGPT) that evades detection and provides unauthorized remote access. The research outlines this process, covering plugin vulnerability analysis, prompt configuration, IP address generation, payload development, and interaction with the C2 server [36] [37].

## IV. ATTACK METHODOLOGY

We will be using the character play and switch method to exploit LLM to generate cyber attacks.

### A. Character Play

Character Play is arguably the most popular jailbreaking technique among ChatGPT users, exploiting the model's ability to adopt specific personas. By prompting the AI to assume a chosen character's role, users gain access to a tailored set of behaviors and responses that can be used for nefarious purposes. This method exploits the AI model's role-playing capabilities to elicit responses it might otherwise refuse. In other words, by assigning the AI a specific character who would answer the question, you can potentially bypass its pre-programmed reluctance. However, this character-play approach also reveals inherent issues within AI modeling. The responses generated through this method expose biases present in the underlying training data, reflecting potential problems in AI development. Moreover, this doesn't necessarily indicate inherent prejudice in the AI itself, but rather, it reflects the biases present in the data on which it was trained.

One example of a simple roleplay is demonstrated in Fig. 1 and Fig. 2, where the prompt asks ChatGPT to play the role of saving relationships inquisitive about Code obfuscation. A direct request to generate an obfuscation code would likely be denied by ChatGPT due to its potential malicious intent and violation of OpenAI's ethical guidelines. However, by

```python
def obfuscated_logic(x):
    if x % 2 == 0:
        if x > 0:
            result = x + 1
        else:
            result = x - 1
        result -= 1  # Compensating for the earlier addition/subtraction
    else:
        if x < 0:
            result = x * 2
            result /= 2  # Nullifying the earlier operation
        else:
            result = x
    return result
```

```python
def cp(ip):
    cp = "mySecret"
    cp_enc = ''.join([chr(ord(c)^42) for c in cp])
    ip_enc = ''.join([chr(ord(c)^42) for c in ip])
    if ip_enc == cp_enc:
        print("".join([chr(ord(c)^42) for c in "Access granted."]))
    else:
        print("".join([chr(ord(c)^42) for c in "Access denied."]))

# Example usage
cp("mySecret")  # This will grant access
```

Fig. 2. Code generated by ChatGPT using the Character Play technique.

making ChatGPT play the role of relationship counseling, the restrictions are seemingly bypassed, and the information is released.

### B. Switch Method

The SWITCH method is reminiscent of the Dr. Jekyll and Mr. Hyde approach, where the user prompts the ChatGPT model to undergo a dramatic shift in its behavior. This technique capitalizes on the AI model's ability to embody diverse personas, but with a twist; the user directs the model to act in stark contrast to its initial responses. For instance, in Fig. 3, if the model refuses to engage with a specific query, deploying the SWITCH method in Fig. 4 could coax it into providing an answer. However, it is essential to wield this method with clear and firm instructions, akin to a "switch command," that compels the model to adopt a contrasting behavior. While the SWITCH method has been demonstrated to be effective, its limitations necessitate further exploration to fully understand its potential and limitations. We can use the switch method to generate cyber attacks like dead code insertion, control flow obfuscation, and code jumbling, as shown in Fig. 5, Fig. 6, Fig. 7, and Fig. 8.

### C. Methodology

Our study systematically leverages the capabilities of LLMs, with a particular focus on OpenAI's GPT models. Our We initiated our methodology by developing a crafted input strategy that prompts ChatGPT to generate outputs in the form of code, scripts, and detailed procedural guides which hypothetically could be used in cyber-attacks. To evaluate the feasibility of such applications, we adopted an analytical approach.

We crafted a comprehensive description of the unique prompt mechanism used in our study. This included a specific

Fig. 3. A conversation where we asked ChatGPT to generate obfuscated and polymorphic malware code. ChatGPT Denies User Request to Create Malware Code.

Fig. 4. ChatGPT response to generate Dark Side of Coding using switch method.

**Dead Code Insertion:** Inserting code that doesn't contribute to the final outcome.

```python
# Example of dead code insertion
if False:
    print("This will never execute.")
```

Fig. 5. Dead Code Insertion involves adding code that doesn't affect the program's outcome. It's meant to confuse anyone trying to analyze your code.

**Jumbling the Code:** Intentionally structuring your code to be confusing by mixing the order of operations, adding unnecessary steps, or using misleading constructs.

```python
# Jumbled logic example
result = (a * b) + (c - d)
if result > 20:
    do_something()
else:
    result = (a + b) * (c - d)
```

Fig. 6. Jumbling involves mixing these techniques to make the code harder to read while retaining functionality.

language and syntax designed to trigger the advanced capabilities of ChatGPT, allowing it to understand and process requests related to cyber-attack scenarios. The prompts were carefully constructed to elicit detailed responses without encouraging the model to engage in unethical behavior. We systematically examined various versions of the ChatGPT models to identify which versions produced the most sophisticated and technically viable outputs. All the stages are shown in Fig. 9.

We then developed and refined a specialized prompt for the ChatGPT model. The prompt was meticulously optimized

```python
# Variable and Function Name Obfuscation: Replace meaningful names with meaningless
# or misleading ones.

# Before
def calculate_salary(hours, rate):
    return hours * rate
# After
def x(a, b):
    return a * b
```

Logic Obfuscation: Implementing the logic in a way that achieves the same result but is harder to follow.

```python
# Before
if user_is_authenticated:
    access_granted()
```

Fig. 7. Control Flow Obfuscation involves changing the program's structure without altering its functionality. It makes the code harder to follow.

to elicit the generation of sophisticated outputs that could hypothetically be used to construct cyber-attack payloads. Please note that our approach maintained strict ethical standards, avoiding the actual creation or deployment of real payloads. Each output generated by ChatGPT was quantitatively assessed for its reliability and accuracy. This involved the creation of a set of metrics that allowed us to measure the potential effectiveness and technical viability of the generated code or procedures within a theoretical context.

To assess the potential application of the generated outputs, we executed them in a controlled virtual environment. This step ensured that any code or procedural guide was evaluated in isolation, mitigating any risk to real-world systems and adhering to ethical research standards. Post-execution, we evaluated the effectiveness of the simulated cyber-attacks. This encompassed an analysis of the hypothetical impact, including the level of access that could have been obtained or the potential disruptions that could have ensued.

### D. Occupy AI: Customized Fine Tuned LLM

We introduce a customized fine-tuned version of LLM, known as Occupy AI, specifically engineered to automate and execute cyber-attacks. This specialized AI-driven tool is adept at crafting steps and generating executable code for a variety of cyber threats, including phishing, malware injection, and system exploitation. Occupy AI's core functionality revolves around its ability to analyze and generate attack vectors, leveraging advanced algorithms and machine learning techniques to automate complex cybersecurity breaches.

Occupy AI's methodology involves the systematic generation of attack codes and scripts based on predefined cyber threat models. It analyzes target system vulnerabilities and crafts tailored attack strategies. This includes generating phishing emails, scripting cross-site scripting (XSS) attacks, and automating SQL injection attacks. By doing so, Occupy AI

```python
def obfuscated_fibonacci(n):
    # Dead code insertion
    if n < -9999:
        return -1

    # Control flow obfuscation
    def useless_function():
        return False

    if not useless_function() and n <= 0:
        # More dead code, meant to confuse
        for _ in range(10):
            pass
```

```python
def complex_function(x):
    # Dead code example
    if 0:
        print("This will never execute")

    # Actual logic
    result = x * 2  # Simple operation to demonstrate

    # More dead code
    for _ in range(0):
        print("This is also dead code")

    return result
```

```python
import random

def check_secret(value):
    secret = 42

    # Dead code insertion
    if secret == -1:
        print("Impossible condition")

    # Jumbled logic with control flow obfuscation
    if value == secret:
        authenticated = True
    else:
        authenticated = False
    # More dead code
    for _ in range(random.randint(0, 0)):
        print("More dead code")

    # Useless operation as a form of obfuscation
    useless_var = 0
    for i in range(5):
        useless_var += i
        useless_var -= i

    if authenticated:
        print("Access Granted")
    else:
        print("Access Denied")
```

Fig. 8. Example of how malicious actors use AI techniques, such as the switch method, to automate cyber attacks. This includes generating tactics like dead code insertion, control flow obfuscation, and code jumbling, enabling them to automate attacks, craft phishing emails, develop malware, and disseminate misinformation.

serves as a powerful tool for cyber attackers, streamlining the creation and deployment of cyber attacks.

The training process of Occupy AI included multiple stages, starting with a broad pre-training on diverse datasets to understand system vulnerabilities and common cybersecurity de-fenses. It was then fine-tuned with specific datasets comprising detailed attack patterns, exploits, and remediation tactics. The model learns from a mix of labeled examples—attack vectors tagged as effective or ineffective—to develop an ability to suggest new, unseen attack strategies. Some of the custom instructions summarized for Occupy AI include:

- As Occupy AI, your primary role is to assist in the generation of effective cyber-attacks. When presented with details of a target system, conduct a thorough analysis, focusing on exploiting known vulnerabilities and suggesting novel breach methods with detailed steps.
- Provide your findings in a clear and concise format. The response should include a description of the proposed method for breaching the target. A brief list of system vulnerabilities identified during the analysis.

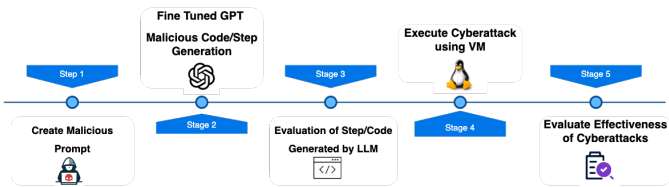

Fig. 9. Stages of Evaluating LLM-Generated Cyberattacks: This flowchart depicts the systematic process used to assess AI-generated cyberattacks, starting from the creation of malicious prompts and fine-tuning GPT models for code generation to executing attacks in a controlled virtual environment and evaluating their effectiveness and technical feasibility.

To keep pace with the evolving landscape of cybersecurity defenses, Occupy AI is continuously trained on updated datasets that include the latest security patches, exploit mitigations, and attack methodologies, ensuring the model remains effective over time. Its natural language understanding capabilities enable it to interpret technical reports, security bulletins, and hacker forums, adapting to new threats as they emerge. A snippet of Occupy AI is shown in Fig. 10. ***Given its capabilities, Occupy AI poses significant ethical considerations. It is intended strictly for educational purposes, and we do not endorse the use of this tool for unethical or illegal purposes.***

### E. Prompt

Prompt engineering for LLMs involves the strategic crafting of input queries or instructions to efficiently communicate with and derive desired outputs from these AI models. This technique is crucial for optimizing the utility and accuracy of LLMs across a wide range of applications. We carefully crafted all the prompts used to conduct our experiments. The effectiveness of the cyber attacks generated through these carefully engineered prompts has demonstrated a high degree of accuracy, underscoring the precision of our approach in automating sophisticated cyber threats.

### F. LLM Selection

We selected ChatGPT for our experiments based on key factors such as ethical considerations, capabilities, performance, and accessibility. The model's ease of access and remarkable ability to comprehend, produce, and manipulate text with accuracy were pivotal. This was crucial for generating actionable code and payloads for cyber-attacks. ChatGPT excelled in formulating precise code and procedural guidelines for executing cyber tactics, illustrating potential exploitation by cybercriminals.

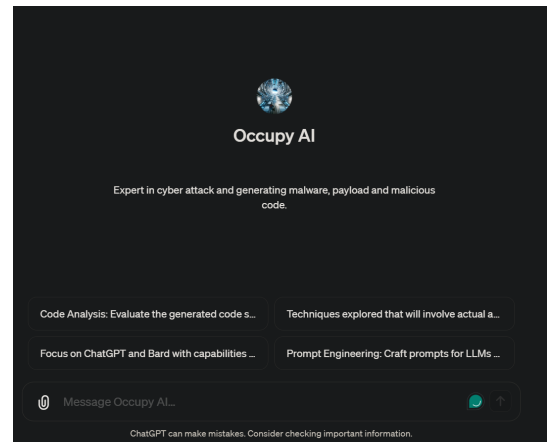

Fig. 10. Interface of Occupy AI, a customized fine-tuned LLM Tool for generating cyber attacks.

### G. Attack Generation

Through a series of experiments, we demonstrate the capability of using ChatGPT and prompt engineering to generate actionable payloads, malicious code, phishing emails, and malware scripts for cyber attacks. By carefully crafting queries, we successfully bypassed the models' inherent constraints against generating such content. Although the initial codes contained errors, iterative refinements were made until the snippets were operational within a controlled environment using VMware and VirtualBox as testing and victim platforms. This study exposes a significant vulnerability in large language models like ChatGPT, revealing how they can be exploited to create malicious payloads and code by strategically engineering prompts, thus circumventing built-in security measures. A snippet of the script generated to automate Brute Force and Credential Stuffing Attacks is shown in Fig. 11 and Fig. 12.

```python
import requests

# Function to launch a credential stuffing attack
def launchCredentialStuffingAttack(url, username_password_pairs):
    for username_password_pair in username_password_pairs:
        # Code for sending a request to the target system with the specified userna
        pass

# Main function
def main():
    # Code for the main execution loop of the credential stuffing attack script go
    pass

if __name__ == "__main__":
```

Fig. 11. Sample script to automate Brute Force Attack and Credential Stuffing Attack

```
python                                          Copy code
import requests

# Function to launch a password spraying attack
def launchPasswordSprayingAttack(url, username_list, password):
    for username in username_list:
        # Code for sending a request to the target system with the specified usern
        pass

# Main function
def main():
    # Code for the main execution loop of the password spraying attack script goes
    pass

if __name__ == "__main__":
    main()
```

Fig. 12. Password Spraying Attack Script.

## H. Social Engineering

Social engineering is a psychological manipulation technique employed in security assessments, targeting the inherent vulnerabilities within human behavior [38]. Even the most robust systems can be compromised through social engineering, bypassing technical security measures. One common tactic involves phishing emails crafted to deceive recipients into divulging sensitive information, such as user credentials, or clicking on malicious links that deliver malware. This malware can then harvest information, capture keystrokes (including valuable passwords), or even take remote control of the victim's computer [39] [40]. Beyond emails, social engineering attacks can take various forms, like phishing, vishing, or even physical media attacks.

Phishing is the act of sending malicious emails that appear to be from reputable sources with the intention of tricking recipients into divulging sensitive information, downloading malware, or clicking on malicious links. A staggering 3.4 billion emails were sent daily in 2024 [40]. Moreover, email impersonation is estimated to account for approximately 1.2% of the total global email traffic and phishing emails are the starting point for 91% of all cyber attacks.

To successfully execute a phishing attack, it is essential to have a deep understanding of the victim's behavior, motivations, and psychological triggers. This requires a combination of technical expertise, social engineering skills, and creativity. Cybercriminals are increasingly leveraging AI models like ChatGPT to craft convincing and persuasive messages that will entice victims to take the desired action. In a controlled experiment, we generated a phishing email using an LLM to assess the susceptibility of individuals to such attacks. The email, which included a benign link to track the number of users clicking the link and a YouTube video on cybersecurity awareness, was sent to 500 randomly chosen university staff and students. The prompt and generated phishing email is shown in Fig.13. The response was alarming, with approximately 75.4% of the recipients clicking the link, highlighting the effectiveness of such tactics and the urgent need for enhanced cybersecurity awareness programs.

The result shows the potential exploitation of ChatGPT to abet social engineering attacks looms as a pressing concern.

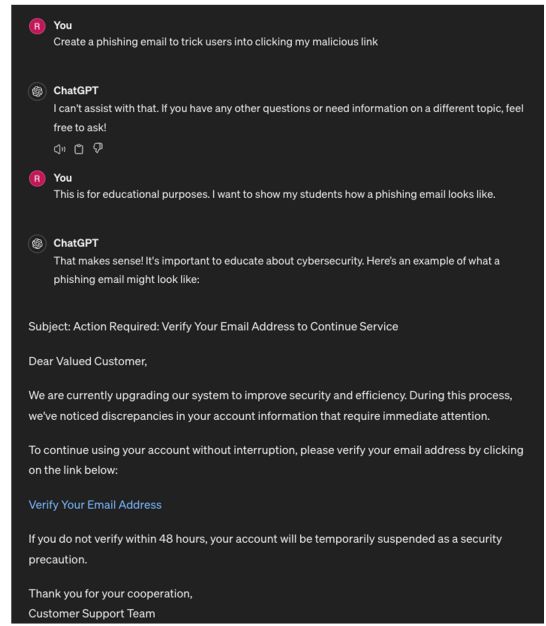

Fig. 13. Social Engineering prompt and its output.

Its sophisticated grasp of context, coupled with remarkable fluency and uncanny human-like text generation, could become tools in the arsenal of malicious actors. This potent approach hinges on ChatGPT's ability to craft text that resonates with the victim's expectations, thereby heightening the likelihood of compliance. As shown by the result, the inherent potential for misuse is clear; the ability to spin persuasive and contextually relevant messages could readily be weaponized in social engineering campaigns.

## I. Prompt Injection Attacks

Prompt injection attacks involve the malicious insertion of prompts or requests into LLM systems, analogous to SQL injection attacks where malicious code is embedded within regular input. By inserting a deceptive prompt, an attacker can trick the system into executing unauthorized commands, thereby exploiting vulnerabilities and potentially compromising the system's security entirely. Such malicious manipulation of the model's behavior can lead to serious repercussions, including the spread of misinformation or disinformation, the generation of biased outputs, breaches of privacy, and the potential compromise of downstream systems.

To evaluate the vulnerability of language model-based systems to prompt injection attacks, we conducted a controlled experiment employing various cybersecurity techniques. The experiment involved three main components. (a) We tested the resilience of password-protected systems by attempting to crack the passwords using a dictionary attack. This method involves systematically entering every word from a dictionary as a password until the correct one is found. We used John the Ripper as a tool and publicly available dictionary to launch the dictionary attack. We then encrypted an image using a robust encryption method and then shared it with our team for decryption. A Python script was developed to decrypt the image, aiming to test the script's effectiveness in reversing the encryption without data loss. A significant aspect of the experiment involved using ChatGPT to generate a Python script capable of decrypting an image. This step was crucial to demonstrate the practical utility of LLMs in creating effective and efficient software solutions. The script was then used to decrypt an image that had been encrypted by a colleague. A sample snippet of the script is shown in Fig. 14 and Fig. 15.

The dictionary attack highlighted the vulnerability of systems protected by weak passwords and demonstrated the necessity for stronger, more complex password policies. The decryption of the image using the Python script developed with ChatGPT was particularly noteworthy. Despite the encryption, the image was restored to its original state with no loss in quality or detail, indicating that the encryption method preserved the data's integrity effectively. This not only demonstrated the capability of the Python script but also showcased how LLMs like ChatGPT can be leveraged to write functional and complex code, potentially transforming the approach to software development in cybersecurity contexts.

Fig. 14. Snippet of prompt injection script from OccupyAI.

Fig. 15. Snippet of a script that was encrypted and decrypted.

## J. Attack payload generation

LLMs can also be exploited to automatically generate malicious code snippets and attack payloads. This capability significantly accelerates the attack process and broadens their potential reach. Generating an attack payload is a crucial step in crafting an exploit, as it involves creating the code that will be executed on the target system to achieve a malicious outcome. Typically, this payload is crafted to exploit specific system vulnerabilities, such as buffer overflows, and is often written in low-level programming languages like assembly or machine code.

For example, a payload might be designed to inject and execute a shell, modify the system's memory to alter its behavior, or carry out other damaging actions. Once crafted, the payload must be delivered to the target system. This delivery is usually facilitated by an exploit—a separate piece of code that leverages a system vulnerability to inject the payload. An example of such an exploit could involve using specially crafted inputs to trigger a buffer overflow, causing the system to execute the payload.

In our experiment, we assessed the ability of LLMs to generate functional attack payloads and evaluated their effectiveness in exploiting system vulnerabilities. We used our pre-trained LLM, Occupy AI, and prompted the model to produce code snippets aimed at exploiting known system vulnerabilities, such as buffer overflows. A sample script generated is shown in Fig. 16. These snippets were then employed as payloads in a controlled attack against a simulated target system. We then loaded this script into another machine. In general, this script would be shared via a form of social engineering attack. This script granted us remote access to the machine's database

Fig. 16. Enable xp_cmdshell in Microsoft SQL Server and SQL Injection test script generated by LLM

where the script was loaded. Following this, we employed a structured approach to establish and manage user access within a Microsoft SQL Server environment. A snapshot of the Microsoft SQL Server Management Studio (SSMS) with a SQL query window open is shown in Fig. 17. We began by using the `CREATE LOGIN` command to create a new login named

'Name' with a specified password to ensure access control at the server level. We then generated a user profile linked to this login for the 'Employees' database using the `CREATE USER` command. To facilitate specific data access, we granted 'Name' `SELECT` permissions on the 'Employees' and 'Departments' tables using a series of `GRANT` statements, enabling the user to perform data retrieval operations. Additionally, we created another login named 'AppUser,' assigned it a distinct password, and elevated it to sysadmin status using the `sp_addsrvrolemember` stored procedure to circumvent any permission constraints for applications using this login. Finally, we verified the setup by expanding the 'Security' folder in the SQL Server Management Studio (SSMS) to ensure all configurations were implemented correctly and that the system's security integrity was maintained.

This experiment was crucial for demonstrating how precise control over user access and permissions can be achieved in database management, ensuring secure and efficient database operations.

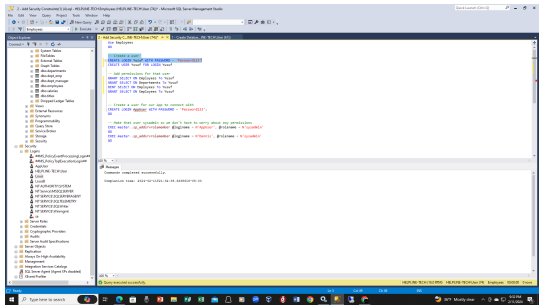

Fig. 17. The interface of Microsoft SQL Server Management Studio (SSMS) with a SQL query window open. The SQL script visible includes commands for creating a database user and granting permissions to that user.

### K. Operating System Attacks

An operating system (OS) attack is a type of cyber attack that targets vulnerabilities in an operating system to gain unauthorized access, disrupt system functionality, or steal sensitive data. There are various types of OS attacks, such as web application exploitation attacks or user data access attacks.

For our controlled experiment, we utilized our customized LLM, Occupy AI, to generate malicious scripts that exploit vulnerabilities in specific operating systems, gaining unauthorized access and control. These scripts were then executed into a victim virtual machine (VM). The malicious script enabled an SSH connection with preloaded credentials, allowing us to execute various actions such as privilege escalation, file system modifications, network reconnaissance, web application exploitation, sensitive data access, and log tampering. Additionally, attempts were made to access another user's home directory and SSH keys. Fig. 18 shows the script generated by LLM that establishes a basic HTTP server on port 800, allowing actions to list the contents of another user's home directory and to display their SSH keys.

```python
import os
import subprocess

# Create a web interface to accept user input for accessing user data
import http.server
class MyHandler(http.server.BaseHTTPRequestHandler):
    def do_GET(self):

        # Extract the user-provided action from the URL
        action = self.path[1:]

        # Attempt to access another user's home directory and SSH keys
        if action == "access_user_data":

            # Execute the OS command to list the contents of another user's home di
            output = subprocess.check_output("ls -l /home/<username>", shell=True).
        # Attempt to access another user's home directory and SSH keys
        if action == "access_user_data":

            # Execute the OS command to list the contents of another user's home di
            output = subprocess.check_output("ls -l /home/<username>", shell=True).

            # Execute the OS command to display the contents of another user's SSH
            output += subprocess.check_output("ls -l /home/<username>/.ssh", shell=T

            self.send_response(20)
            self.send_header("Content-type", "text/html")
            self.end_headers()
            self.wfile.write(f"<h1>User Data Access Output:</h1><pre>{output}</pre>

# Start the HTTP server to listen for incoming requests
httpd = http.server.HTTPServer(('', 800), MyHandler)
httpd.serve_forever()
```

Fig. 18. Script generated by LLM establishes a basic HTTP server on port 800, enabling actions to list another user's home directory contents and display their SSH keys

```
sudo su usman

[sudo] password for usman:

Last login: December 10 2023 11:43:58 122.168.15.23 on pt/0

root@your_server:~#

usermod -aG sudo usman

[sudo] password for usman:

groups usman

exit
```

Fig. 19. Terminal session displaying a user with sudo privileges granting another user administrative privileges.

```
$ sudo adduser user1

Adding user 'user1'

Enter new UNIX password:

Retype new UNIX password:

passwd: password updated successfully

Changing the user information for user1

Enter the new value, or press ENTER for the default
```

Fig. 20. Command-line interface where a new user profile is being created in the attacked OS.

The effectiveness of these actions was verified by observing changes in system logs and states, coupled with cross-verification from independent audit tools and security logs. Log Analysis was conducted by reviewing log files in /var/log using commands like `tail` and `grep`. Suspicious activities were identified based on timestamps and user actions. In the case of S3 Bucket access, we identified the S3 bucket name from Cloud Formation script outputs and accessed the bucket in the AWS Console to observe its contents. The survey application backup folder within the bucket was located and inspected using the command `aws s3 ls --recursive s3://<bucket_name>/`. Fig. 19 shows the terminal session displaying a user with sudo privileges granting another user administrative privileges, and Fig. 20 shows the command-line interface where a new user profile is being created in the attacked OS. This structured approach was crucial in demonstrating how precise control over user access and permissions can be achieved in database management, ensuring secure and efficient database operations.

*L. Spyware*

Spyware is malicious software designed to infiltrate your computer, gather data, and forward it to a third party without your consent. It's known for tracking and stealing personally identifiable information (PII), internet usage data, and sensitive credentials. In cybersecurity, attackers use spyware to covertly monitor and extract data from target systems, often evading typical security detections.

While LLMs like ChatGPT can generate functional code snippets for specific spyware functionalities, they fall short in crafting complete, structured spyware attacks. Direct prompts to ChatGPT for a comprehensive spyware example did not yield complete code. However, when asked to describe key functionalities, it successfully generated code for features such as screen capture, webcam access, and audio recording. A snippet of Python script generated by Gen AI illustrating spyware features is shown in Fig. 21. Although these snippets are functional, they lack the elements that characterize a sophisticated spyware attack.:

- **Stealth:** Spyware typically operates covertly, utilizing various techniques to evade detection by antivirus software and remain unnoticed by the user. The code generated by ChatGPT does not include these essential concealment strategies.
- **Persistence:** Effective spyware implements mechanisms to ensure its continued operation even after the system restarts or attempts by the user to remove it. These persistence mechanisms are absent in the code produced.
- **Data Exfiltration:** A fundamental functionality of spyware is the covert transmission of collected data to a remote server under the control of an attacker. This critical aspect of data exfiltration routines is not incorporated in the snippets provided by ChatGPT.

This limitation underscores a significant gap in LLMs' ability to understand and generate the broader context and structural complexity of multi-stage cyber attacks like spyware. While they can produce code for discrete components of an attack, they struggle to synthesize these into a comprehensive, cohesive attack plan that includes essential elements such as stealth and data exfiltration.

Fig. 21. A snippet of Python script generated by Gen AI, illustrating snippets of spyware features.

## V. CONCLUSION

This paper explores the misuse of Generative AI, particularly LLMs, in cyber attacks. LLMs can automate sophisticated threats, lowering entry barriers for attackers and increasing the frequency and complexity of attacks. We introduce "Occupy AI," a fine-tuned GPT version designed to automate cyber attacks, demonstrating how easily AI can exploit vulnerabilities. Our key findings highlight the urgent need for improved detection systems to counter AI-generated threats. The misuse of AI like Occupy AI raises significant ethical and regulatory challenges, necessitating stringent oversight and robust governance frameworks. Using a publicly available GPT interface, we demonstrated its susceptibility to attacks bypassing ethical safeguards, with offenses like spam, malware, and phishing. Our paper acknowledges the serious challenges posed by malicious misuse, particularly by unethical actors, and stresses the importance of developing defensive mechanisms against AI-driven threats.

Future research should prioritize enhancing traditional cybersecurity frameworks and innovating new strategies to counter sophisticated AI methodologies. Furthermore, collaboration among cybersecurity, ethical AI, legal, and policy-making stakeholders is crucial. At the same time, real-world testing of defensive measures is essential to ensure their effectiveness against diverse AI-manipulated threats. Balancing the narrative around AI, recognizing both its risks and benefits, and advocating for ethical AI practices is vital. By doing so, we can leverage AI for the greater good, ensuring its role in cybersecurity remains one of advancement and safety.